\documentclass[12pt,preprint]{aastex}

\slugcomment{}

\shorttitle{Hydrodynamic Instability of Ionization Fronts}
\shortauthors{Mizuta et al.}

\begin{document}


\title{Hydrodynamic instability of ionization fronts in HII regions}


\author{Akira Mizuta\altaffilmark{1,2,3}, Jave O. Kane\altaffilmark{3},
Marc W. Pound\altaffilmark{4},
Bruce A. Remington\altaffilmark{3},\\
Dmitri D. Ryutov\altaffilmark{3},
and Hideaki Takabe\altaffilmark{1}}


\altaffiltext{1}{Institute of Laser Engineering Osaka University,
2-6 Yamada Oka, Suita, Osaka, 565-0871, JAPAN}
\altaffiltext{2}{amizuta@ile.osaka-u.ac.jp}
\altaffiltext{3}{University of California, Lawrence Livermore National Laboratory, 7000 East Ave., Livermore, CA 94551}
\altaffiltext{4}{Department of Astronomy, University of Maryland, College Park, MD 20742}


\begin{abstract}
We investigate hydrodynamic instability of accelerating
ionization fronts
using two dimensional hydrodynamic simulations
that include detailed energy deposition and release
due to the absorption of UV radiation,
recombination of hydrogen,
radiative molecular cooling, and magnetic pressure.
We consider linear perturbation growth and find that
the stabilization mechanism associated with non-accelerated fronts
remains a significant factor even when acceleration is present.
Conversely, if recombination in the ionized region is turned
off, Rayleigh-Taylor (RT) instability becomes effective,
and the classical RT growth rate recovered.
\end{abstract}


\keywords{HII regions --  ISM: molecules --  ISM: kinematics and
dynamics -- hydrodynamics -- instabilities -- methods: numerical}


\section{INTRODUCTION}
Ionization fronts (IF) near OB stars
have attracted considerable interest due to their amazing shapes.
For example, the Eagle Nebula has
three famous `pillars' or `elephant trunks'
\citep{Hester96,Pound98}.
Strong UV radiation from nearby OB stars illuminates
the surface of a molecular cloud,
and photoevaporation occurs in a very thin layer at the surface.
As a result, an ablation flow begins
and a shock propagates into the cloud.
Such an ionization or ablation front is categorized as D-type.
The size of the system including the OB stars and the molecular cloud
is a few parsecs, and
the width of pillars is usually less than a parsec.

A number of previous theoretical works and numerical simulations
(\citet{Williams01}, \citet{Williams02} and references therein)
have addressed understand the formation of pillars,
but this problem is still an area of active research.
All the models considered thus far can be divided into two large
categories:
those that relate formation of pillars to the presence of
pre-existing dense clumps `excavated' by moving ablation front,
and  those that relate the pillars to a nonlinear stages of
ionization front instabilities.
In this article we study the second category.
The first category has been also extensively studied  by various authors,
most recently by \citet{Williams01},
who reference earlier work. 

In modeling ionization front instability,
researchers have pursued one of two approaches:
they have either studied instability of the RT type,
associated with finite acceleration of the cloud driven
by the ablation pressure, or studied instability of
a non-accelerating front associated with details of
ionization/recombination processes in the ionized outflow.
As far as we know, our paper
will be the first where both the acceleration and
ionization/recombination processes
are taken into account simultaneously.

To put our paper into historical perspective,
we mention that RT instability of molecular clouds
was studied by \citet{Spitzer54} and \citet{Frieman54}.
The case where the incident radiation is tilted with respect to the cloud surface
was studied in \citet{Ryutov03}.
Instability of non-accelerated ionization fronts was
studied by \citet{Vandervoort62}, with the radiation tilt included.
\citet{Kahn58} argued qualitatively that absorption
of the incoming radiation by the hydrogen atoms formed
by recombination in the ablation outflow
can have a stabilizing effect,
because of the stronger absorption near the dimples of
the surface relief compared to the bumps.
\citet{Axford64} presented a quantitative study that showed
that this stabilization mechanism is most effective
for perturbations with wavelengths larger than the recombination length.
\citet{Sysoev97} provided more complete analysis and
find the growth of long wavelength instabilities for normally
incident radiation.
\citet{Williams02} confirmed it and included effects
of the radiation tilt.

The main result of our study is that the linear RT instability,
which is usually very robust and hard to stabilize,
is actually also stabilized by the recombination effect. 

\section{BASIC EQUATIONS}
Most theoretical analyses and numerical simulations
assume that the gas is isothermal and
do not take into account the energy budget.
However it is possible to formulate the problem including detailed energy
treatment.
At the ablation front,
an incident photon whose energy is higher than
the Lyman limit is absorbed, ionizing the neutral gas
and increasing its thermal energy.
In the ionized region
ionized and electrons recombine
and thermal energy is radiated .
When an ablatively driven shock propagates into the molecular gas,
the gas is radiatively
cooled on a very short time scale \citep{Neufeld95}.
The temperature of the interior of the molecular cloud in the Eagle Nebula
pillars is a few tens of Kelvins \citep{Pound98}.
In our work we solve the equations of two dimensional hydrodynamics with
energy source terms that  account for the radiative and
recombinative processes just described:
\begin{eqnarray}
{\partial \rho \over \partial t}+
\nabla\cdot (\rho\mbox{\boldmath$u$})=
0,\\
{\partial (\rho\mbox{\boldmath$u$})\over \partial t}+
\nabla\cdot (\rho\mbox{\boldmath$u$}\mbox{\boldmath$u$}+p\mbox{\boldmath$I$})=
0,\\
{\partial \over \partial t}\left(\rho\left({1\over 2}\mbox{\boldmath$u$}^2
+\epsilon\right)\right)+
\nabla\cdot 
\left(
\left(\rho\left({1\over 2}\mbox{\boldmath$u$}^2
+\epsilon\right)+p\right)\mbox{\boldmath$u$}
\right)=-q_{re}+q_{uv}-q_{mol}.
\end{eqnarray}
In order, these are the mass, momentum and energy conservation equations.
$\rho$ is mass density, $p$ is pressure,
$\mbox{\boldmath$I$}$ is unit tensor,
\mbox{\boldmath$u$} is the velocity vector,
and $\epsilon$ is the specific internal energy.
$q_{re}$, $q_{uv}$, and $q_{mol}$ are
energy source terms due to the recombination in the ionized region,
absorption of UV radiation from OB stars, and
cooling in the molecular gas due to rotation and
vibration mode of hydorgen molecules, respectively.

Photoionization and recombination are considered with
the following equations:
\begin{eqnarray}
\label{fraction}
n{\partial f\over \partial t}+
n \mbox{\boldmath$u$}\cdot\mbox{\boldmath$\bigtriangledown$}f=
an(1-f)J-\alpha_{B}n^2f^2,\\
{\partial J\over \partial y}=-an(1-f)J,
\end{eqnarray}
where $f=n_{i}/n$ is the ionization fraction
($f=0$ or 1 correspond to neutral or fully ionized gas, respectively),
$n$ is the total number of hydrogen nuclei per unit volume,
including nuclei in molecules, atoms and ions,
and $n_{i}$ is ionized hydrogen volume density.
For numerically simplify,
we do not take into account for the state of the neutral
hydrogen,
assuming that all molecular hydrogen dissociates
before the IF,
although the photodissociation region dominates some
thickness \citep{Tielens85} which is longer than that
of photoionization front.
The temperature in the HI region becomes about a few tens to
a few hundreds Kelvin.
Our treatment is just as same as the isothermal models done in a number of
theoretical and numerical works so far.
We also do not consider the energy budget of the dissociation process.
$a=6\times 10^{-18}\mbox{cm}^{2}$ is photoionization cross-section of hydrogen,
and $J$ is the number flux of ionizing photons whose energy exceeds
the critical energy for the ionization of hydrogen (13.6eV).
We assume that the incident photons are parallel to the y axis.
We do not include recombination to the ground state,
instead assuming diffuse radiation and its absorption are balanced locally
(on the spot approximation),
since the effect of this diffuse photon radiation is estimated at $10\% \sim
20\%$ \citep{Canto98}.
Only the so-called ``case B'' recombination
($\alpha _B=2.6\times10^{-13}\mbox{cm}^{3}\mbox{s}^{-1}$ at $T=10^4$K
\citep{Hummer63})
is considered and fixed in this study,
where $\alpha _B$ is the sum of the recombination coefficients
to all hydrogen states except to the ground state. 
Energy sources are written as:
$q_{re}=(nf)^2\beta_{B}k_{B}T$,
$q_{uv}=Wna(1-f)J$,
$q_{mol}=n_{mol}^2\times 
10 ^{-29}\mbox{erg cm}^{-3}\mbox{s}^{-1}$,
and $T=(m_{p}/k_{b})\times[4\epsilon/ (7f+5)]$,
$n_{mol}=n(1-f)/2$,
where $T$ is temperature in Kelvin
($\rho \epsilon =2\times1.5 (nf)k_{B}T+2.5n_{mol}k_{B}T$),
$m_p$ is the proton mass,
$k_b$ is the Boltzmann constant,
and $n_{mod}$ is molecular hydrogen volume density.
We use $(nf)^2\beta_{B}k_{b}T$ instead of $(3/2)(nf)^2\alpha_{B}k_{b}T$ for
the recombination cooling term to model
the thermal velocity dependence of the rates of recombination
and free-free collisional cooling.
$\beta_{B}=1.25\alpha_{B}$ at $T=10^4$ K \citep{Hummer63} is also fixed.
It is assumed that the averaged energy of the incident photon is
$(13.6+W)$ eV per photon from O stars.
The energy $W$ is deposited into the gas as internal energy
when a neutral hydrogen atom absorbs an incident photon.
The ionized region becomes isothermal quickly
due to the energy balance between cooling and heating
processes.
We take $W=1.73 \times 10^{-12}$ erg to produce an isothermal temperature
of $T=10^4$ K in ionized gas.
The cooling term $q_{mol}$ is effective for $40 \mbox{K}< T <
3000 \mbox{K}$.
Since this cooling effect is enough strong to keep the temperature
in the molecular gas 40 K,
this treatment is very similar to isothermal model.
The more realistic treatment was done by \citet{Richling00}
for the calculations of the photoevaporation of protostellar disks. 
The main difference is the introduction of a magnetic pressure
described soon.
We ignore the metal line cooling for simplicity,
since the cooling power has same dependence on
the recombination cooling power ($\propto (nf)^2$)
assuming an equilibrium temperature.
Our conclusion mainly does not change,
whether we take into account the metal line cooling or not.
The temperature $T$ in the ionized outflow can be considered as a free
parameter as it is controlled by the parameter $W$.
In this respect,
an explicit accounting for the radiative cooling does not
seem to be critical to our analysis.
We should note that the heating function $W$ depends on the
temperature of the surface of OB stars.

To close the equations
we also solve the equation of state:
\begin{eqnarray}
p={2(3f+1)\over 7f+5}\rho \epsilon +p_{_{M}}
\left({\rho\over \rho_{_{M}}}\right)
^{\gamma_{_{M}}}.
\end{eqnarray}
The first term on the right hand side is thermal pressure
($f=0$ or 1 give $(2/5)\rho \epsilon$ or $(2/3)\rho \epsilon$
corresponding to  diatomic or monoatomic adiabatic gas)
and the second one is magnetic pressure.
$p_{_{M}}$ and $\rho_{_{M}}$ are constant values.
The index $\gamma_{M}$ ($=4/3 $ in this study) is also constant.
$\gamma_{_{M}}$ can be 4/3 for magnetic turbulence
or 2 for an initially uniform magnetic field.
This magnetic pressure is introduced to prevent radiative collapse
due to molecular cooling when the shock heating occurs \citep{Ryutov02}.
When the gas is compressed,
the second term acts as a stiffness.

The sound speed is then
\begin{eqnarray}
c_{s}^2=
\left({\partial p\over \partial \rho}\right)_{\epsilon}+
{p\over \rho^2}
\left({\partial p\over \partial \epsilon}\right)_{\rho}
=
{2(3f+1)\over 7f+5}\left({p\over \rho}+\epsilon\right)+
{\gamma_{_{M}}p_{_{M}}\over \rho}\left({\rho\over \rho_{_{M}}}\right)
^{\gamma_{_{M}}}.
\end{eqnarray}

The hydrodynamic equations are solved with a Godunov-type code.
The numerical scheme is
the same as used in \citet{Mizuta02},
extended to accept real gas equation of state \citep{Glaister99}.
Photoionization and recombination in Eq. 4 and 5 are solved
implicitly at both half and full step \citep{Williams99}.

\section{NUMERICAL CONDITIONS}
We use two dimensional plane geometry ($x$-$y$).
The grid size is uniform ($\Delta x =\Delta y=2.5\times 10^{-3}\mbox{pc} 
\thickapprox
7.5 \times 10^{15} \mbox{cm}$ which is still longer than the mean free path
of an incident photon for the molecular cloud).

Initially a 0.25 pc thick cloud is set 0.5pc away from the boundary
($y=3$ pc)
where the incident photons are introduced.
The cloud has
initial hydrogen number density n(H)$=10^{5}\mbox{cm}^{{-3}}$ and
temperature $T=40$ K.
Ionized gas is put between the boundary and the cloud surface.
The number density is n(H)$=10 \mbox{cm}^{{-3}}$ and
the gas is pressure matched with the molecular cloud.
Behind the cloud a dilute molecular gas is put (n(H)$=10 \mbox{cm}^{{-3}}$
and
$T=40$ K).
Then this rear cloud surface is not pressure matched initially.
Constant parameters $p_{_{M}}$ and $\rho_{_{M}}$
introduced in our EOS are chosen to be the
thermal pressure and mass density, respectively, of the initial molecular cloud.
The photon flux is parallel to the y-axis and constant except
for a short imprinting period.
All of the gases are at rest at $t=0$
when the radiative flux is turned on.

The dynamics is divided into three phases.
At first, an ablation flow begins
and a shock propagates into the molecular cloud (I).
Since the gas is assumed to be at rest initially,
the ablation front moves in this frame.
This is more realistic than simulating
in the frame where the ablation front is at rest
because we can include the effect of 
increasing distance of the cloud surface from the radiation source.
This effect becomes important in the case with recombination.
The velocity of the ablation front is almost constant in phase I.
When the shock breaks out from the back side of the cloud,
a rarefaction appears and proceeds to the ablation front (II).
This phase is very short, less than 10ky in most cases,
although its length depends on incident photon intensity, number density of the cloud,
and cloud thickness.
When the rarefaction arrives at the ablation front,
the acceleration phase begins (III).

A perturbation is introduced into the incident photon number flux for 10 ky
when phase III begins.
The perturbation is sinusoidal with a 10\% amplitude
$J=J_{0}(1+0.1\cos(2\pi x/\lambda))$, where $J_{0}$ is
the unperturbed incident photon number flux and $\lambda$ is the wave length.
We studied three cases with different wave lengths:
$\lambda =0.92,0.6,$ and $0.46$ pc,
which are similar to the width of the pillars
of the Eagle nebula.
We employ periodic boundary conditions at
both edges that are parallel to y the axis,
and free boundary conditions at the other boundaries.
We follow the evolution at least until the ablation front propagates
a distance equal to the wavelength of the perturbation.
This means the distance between the cloud surface and
the boundary where the photon flux is incident
becomes longer than 1.5 pc in later times.

\section{RESULTS AND DISCUSSION}
\subsection{Without Recombination}
\begin{figure}
\resizebox{12cm}{!}{\plotone{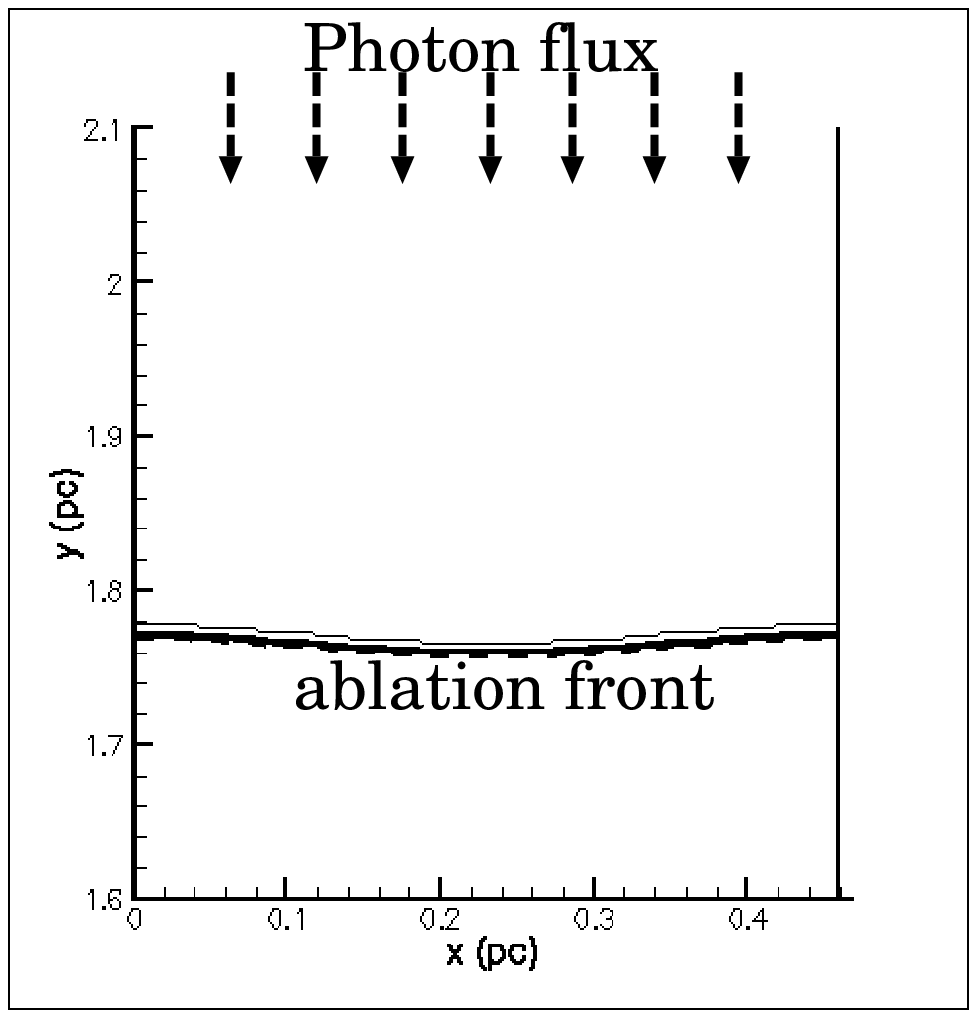}}
\caption{Photon number flux contour (solid lines),
and ablation front ($f=0.5$) contour (dashed line) at $t=340$ky
(without recombination case).
Incident photon flux is incoming from the top of the figure (at y=3 pc).
Photon number flux contours are
drawn every $1.3 \times 10^{8} \mbox{cm}^{-2}\mbox{s}^{-1}$ 
in linear scale at equal intervals.
However, most contours override each other
because of the very short mean free path
of the incident photon in the molecular cloud.}
\label{fluxworecom}
\end{figure}

\begin{figure}
\resizebox{12cm}{!}{\plotone{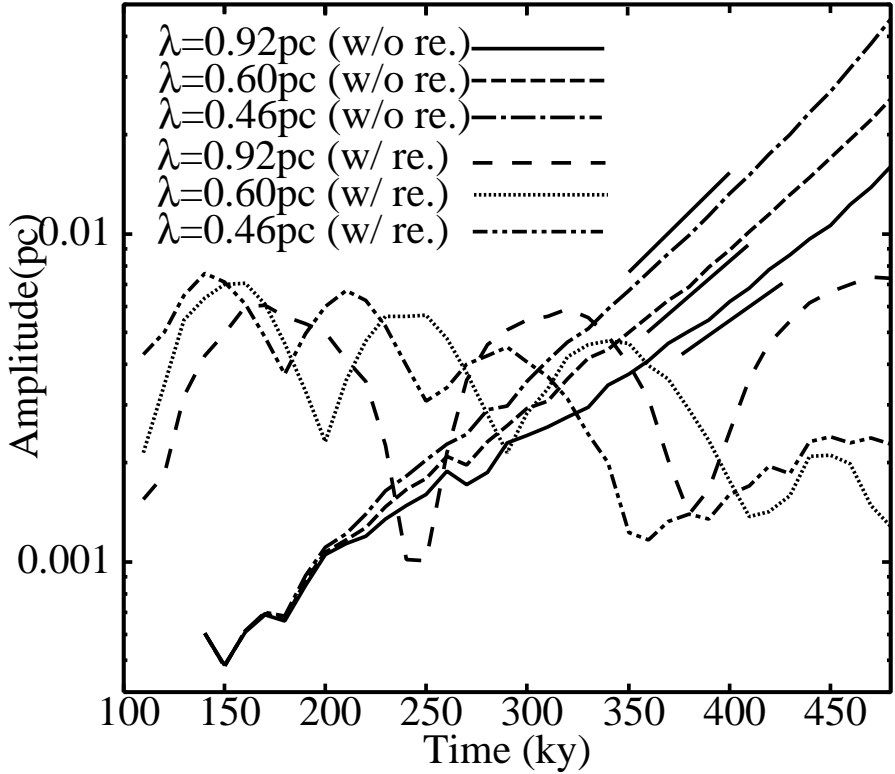}}
\caption{Time evolution of log scale amplitude
of perturbation in each wavelength
(without recombination).
Three wavelengths are shown:
$\lambda=0.92$ pc (solid), $\lambda=0.6$ pc (dotted),
$\lambda=0.46$ pc (dot dash).
The short wavelengths grow faster than the long wavelengths, in accordance
 with classical RT theory.
The short line segments shown are used for the comparison
to  classical RT growth.
Results with recombination cases are also shown:
$\lambda=0.92$ pc (long dased), $\lambda=0.6$ pc (short dashed),
$\lambda=0.46$ pc (two-dot dash).
The shorter wavelengths appear to produce damped oscillation,
whereas the longest wavelength produces oscillation.}
\label{ampworecom}
\end{figure}

At first, we discuss the case without recombination,
namely, $\alpha_{B}=\beta_{B}=0$.
The incident photon flux at the boundary is
$|J_{0}|=2.6\times 10^{9}\mbox{cm}^{-2}\mbox{s}^{-1}$.
Figure \ref{fluxworecom} shows the incident photon number flux (solid lines)
and the ($f=0.5$) ablation front contour at $t=340$ ky.
The absorption occurs in a very thin layer at the ablation
front
because of the very short photon mean free path
because there is no absorption in the ionized gas.
Figure \ref{ampworecom} shows the time evolution of
the perturbation amplitude,
where the amplitude is defined as 
the half distance between the maximum and minimum points in $y$
of the contour $f=0.5$.
Results are shown for the three wavelengths.
The perturbation grows with time for all wavelengths,
and the growth rate is in good agreement with classical RT
theory \citep{Chandrasekhar81}
(the amplitude is
$A=A_{0}\exp(\gamma t)$, where $\gamma=(kg)^{1/2}$ is the growth rate,
$k=2\pi/\lambda$ is the wave number, and $g$ is the effective gravity,
determined from a one dimensional simulation without any perturbation.
The atwood number is assumed to be unity.)
In Figure 2, three short segments shown shown used in the comparison to
RT theory.
Without recombination, the very small imprinted perturbation
(comparable to the grid size),
produces non-linear growth at later times.

\subsection{With Recombination}
On the contrary,
in the case with recombination,
the behavior of the imprinted perturbation is quite different.
The incident photon number flux is increased to
$|J_{0}|=5\times 10^{11}\mbox{cm}^{-2}\mbox{s}^{-1}$
so that the effective gravity in the acceleration phase is almost the same as
in the case without recombination.
This effective flux is possible for the Eagle Nebula,
where the total number flux from O stars is $2\times10^{50} \mbox{s}^{-1}$
\citep{Hester96} and the temperature near the IF is $10^{4}$ K,
although it depends on the distance from the radiation source,
and on the density structure and temperature of the ionized gas.
With this intensity and initial molecular cloud density,
the number density of the ionized gas at the ablation front
becomes about $10^3 \mbox{cm}^{-3}$.
The hydrogen number density of shocked molecular gas is
$\sim 3\times 10^5\mbox{cm}^{-3}$.
These densities are consistent with observations \citep{Hester96,Pound98}.
Figure \ref{ampworecom} shows
time evolution of the amplitude for the three wavelengths,
with recombination.
The number density of ionized gas at the ablation front
decreases when the photon number flux decreases.
The perturbation does not grow,
contrary to the case without recombination,
and the amplitude is oscillating in time.

\begin{figure}
\resizebox{12cm}{!}{\plotone{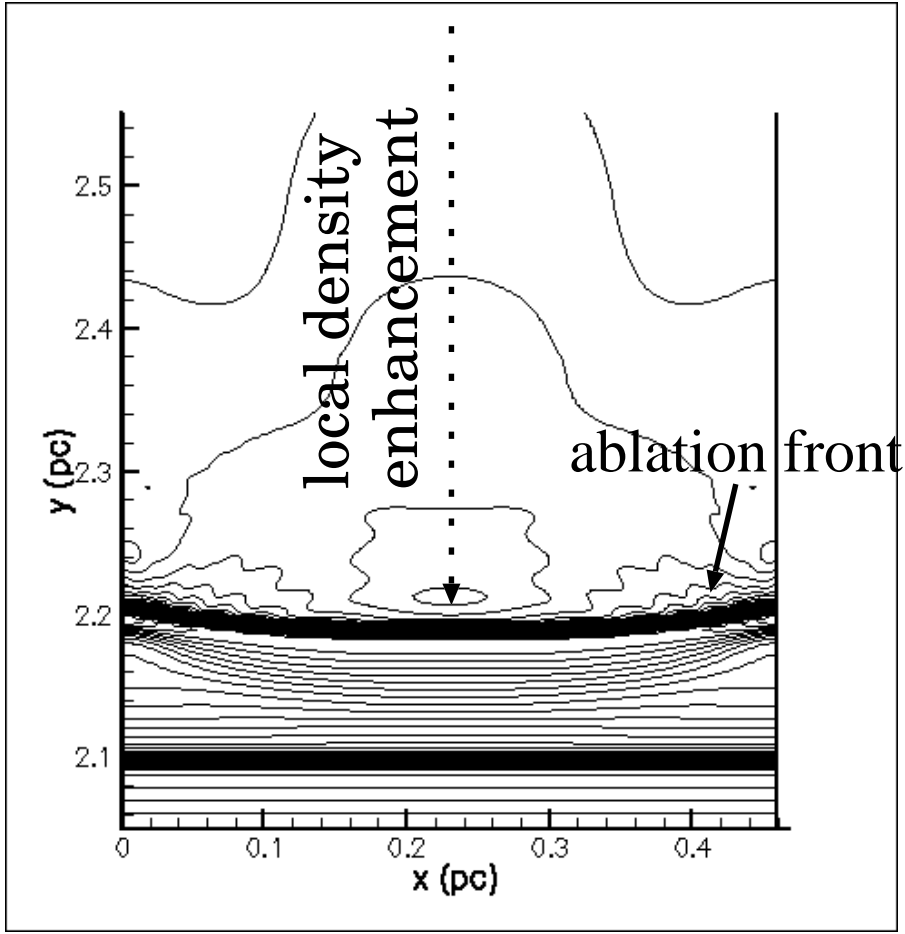}}
\caption{Log scale of hydrogen number density contours at $t=150$ ky
for the simulation including recombination.
Contours are drawn every 0.05 in log scale at equal intervals.
Number density decreases away from the ablation front (see Fig.5).}
\label{rhowrecom}
\end{figure}

\begin{figure}
\resizebox{12cm}{!}{\includegraphics{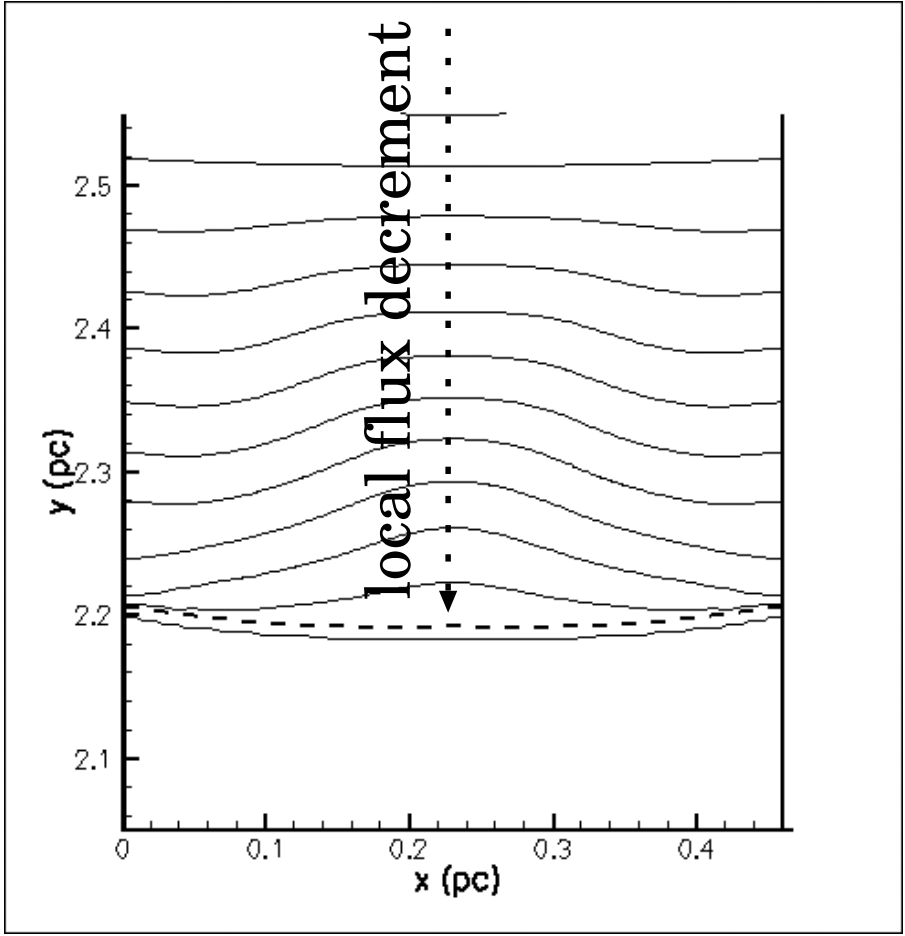}}
\caption{Photon number flux and ablation front contours
(similar to Fig.1) for simulations including recombination
at the same time as the results shown in Fig.4.
Contours are drawn every $2.5 \times 10^{10} \mbox{ cm}^{-2}\mbox{s}^{-1}$ 
in linear scale at equal intervals.
Flux decreases from top to bottom.}
\label{fluxwrecom}
\end{figure}

The qualitative explanation of the absence of the growth
in this case is very similar to that offered
by \citet{Kahn58} and \citet{Axford64}. 
Some of the ablated material in the bubble (concave) region converges
and a dense region appears in the ionized gas near the ablation front
(Fig. \ref{rhowrecom}).
This density structure affects the absorption near the ablation front
(Fig. \ref{fluxwrecom}).
In the dense gas the recombination rate,
which has a quadratic dependence on density (Eq. \ref{fraction}) is high,
and the absorption of incident photons is enhanced.
The photon flux is larger around the spike,
which is strongly driven so that the rocket effect is strong there.
On the other hand, incident photons are absorbed
only moderately around the bubble, and the rocket effect is weaker there.
As a result, the amplitude is reduced.
Once the amplitude is reduced multi-mode patterns appear and
oscillate but do not grow, contrary to the case without recombination.

This large difference between the cases with and without recombination 
results from a competition between
RT push (wanting to make perturbations grow), and
recombination/opacity damping
(tending to make perturbations anneal and flatten out).
Only the RT push works without recombination.
On the other hand with recombination both effects
are available.
Then, a small amplitude perturbation does not grow
because the damping works effectively.
This suggests there may be a critical amplitude
for the RT push to win and produce nonlinear growth.
The growth rate is only weakly affected
by ablative stabilization
(for example \citet{Takabe85}),
because the wavelengths of interest are quite large;
as we consider only normal incidence of the radiation,
effects of radiation tilt \citet{Ryutov03}
are also absent.

We also studied the same problem using
Williams's isothermal model \citet{Williams99} for comparison.
He assumed that both the ionized and molecular gases are
isothermal instead of including a detailed  energy budget
accounting for radiative processes.
The isothermal sound velocity is defined as
$c_{s}=(1+99f)^{1/2} \times 1 \mbox{ km s}^{-1}$.
The pressure and specific internal energy are reset
using this sound velocity and an adiabatic equation of state.
The adiabatic index is assumed to be $\Gamma =1.1$.
The results are very similar to what we have shown here both without and with
recombination,
namely, the perturbation grows without recombination
and does not grow and oscillates with recombination.
Hence, this stabilization effect in the linear regime
is observed using two different methods of simulating
the radiatively driven cloud.

\section{CONCLUSION}
Instabilities at ionization fronts are discussed
with effective gravitation using parameters seen in the Eagle Nebula.
The acceleration phase has not been studied well
and may occur in the evolution of Nebula.
The perturbation on the ionization front
grows in the case without recombination.
The growth rate is in good agreement with classical RT
instability.
When recombination is turned on, which is the more realistic case,
the difference of density profile
causes a different absorption profile.
This works to effectively smooth the surface.
The perturbation does not grow
and the amplitude oscillates in time.
This damping effect works because with recombination
the ablated plasma is not optically thin. 
We observed the same results using Williams' (1999) isothermal model.
Linear regime perturbations appear to exhibit a stabilization effect in these IF simulations.
In this study, the imprinted perturbation is
very small and we observed only the liner growth regime
with recombination.
The dynamics may be quite different in the nonlinear regime.
There may be a critical angle of the surface perturbation above which
the RT push wins out in the competition between RT push and
recombination/opacity damping.

\section*{ACKNOWLEDGMENTS}
We would like to thank R. Williams
for helpful discussions.
This work was performed under the auspices of the U.S.
Department of Energy by the Lawrence Livermore National
Laboratory under Contract No. W-7405-ENG-48 and
NASA Grant NRA 00-01-ATP-059.
One of authors (A.M.) would like to acknowledge the support
from the Japan Society for the Promotion of
Science (JSPS).

\clearpage

\end{document}